\documentclass[aps,prl,twocolumn,notitlepage,showpacs,superscriptaddress,am]{revtex4-1}%
\usepackage{graphicx,ulem}
\usepackage{amsmath,braket}
\usepackage{amssymb}
\usepackage{siunitx}
\usepackage{soul}
\usepackage{color}
\usepackage{amsfonts}%
\providecommand{\U}[1]{\protect\rule{.1in}{.1in}}
\usepackage[svgnames]{xcolor}
\usepackage{changes}
\newif\ifshowcomments
\showcommentstrue

\usepackage{changes}

\def\sro{Sr$_2$RuO$_4\,$}
\def\O{$^{17}$O}
\def\a{$\mathbf{a}$}
\def\b{$\mathbf{b}$}

\def\BBzero{$\mathbf{B_0}$}

\newcommand{\HKS}{\hat{H}_\mathrm{KS}}
\newcommand{\HQP}{\hat{H}^{qp}_{0}}
\newcommand{\TFL}{T_{\mathrm{FL}}}
\newcommand{\chiqp}{\chi^{qp}}

\begin{document}
\title{Tuning the Fermi Liquid Crossover in \sro with Uniaxial Stress}
\author{A. Chronister}
\affiliation{Department of Physics and Astronomy, UCLA, Los Angeles, California 90095, USA}
\author{M. Zingl}
\affiliation{Center for Computational Quantum Physics, Flatiron Institute, 162 5th Avenue, New York, NY 10010, USA}
\author{A. Pustogow}
\affiliation{Department of Physics and Astronomy, UCLA, Los Angeles, California 90095, USA}
\affiliation{Institute of Solid State Physics, TU Wien, 1040 Vienna, Austria}
\author{Yongkang Luo}
\affiliation{Department of Physics and Astronomy, UCLA, Los Angeles, California 90095, USA}
\author{D. A. Sokolov}
\affiliation{Max Planck Institute for Chemical Physics of Solids, Dresden, Germany}
\author{F. Jerzembeck}
\affiliation{Max Planck Institute for Chemical Physics of Solids, Dresden, Germany}
\author{N. Kikugawa}
\affiliation{National Institute for Materials Science, Tsukuba 305-0003, Japan}
\author{C. W. Hicks}
\affiliation{Max Planck Institute for Chemical Physics of Solids, Dresden, Germany}
\author{J. Mravlje}
\affiliation{Jo\v{z}ef Stefan Institute, Jamova 39, SI-1000, Ljubljana, Slovenia}
\author{E. D. Bauer}
\affiliation{Los Alamos National Laboratory, Los Alamos, New Mexico, USA}
\author{A. P. Mackenzie}
\affiliation{Max Planck Institute for Chemical Physics of Solids, Dresden, Germany}
\affiliation{SUPA, School of Physics and Astronomy, University of St. Andrews KY16 9SS, UK}
\author{A. Georges}
\affiliation{Center for Computational Quantum Physics, Flatiron Institute, 162 5th Avenue, New York, NY 10010, USA}
\affiliation{Coll\`ege de France, 11 place Marcelin Berthelot, 75005, Paris, France}
\affiliation{Centre de Physique Théorique Ecole Polytechnique, CNRS, Université Paris-Saclay, 91128, Palaiseau, France}
\affiliation{DQMP, Universit{\'e} de Gen{\`e}ve, 24 quai Ernest Ansermet, CH-1211 Gen{\`e}ve, Suisse}
\author{S. E. Brown}
\affiliation{Department of Physics and Astronomy, UCLA, Los Angeles, California 90095, USA}

\date{\today}

\begin{abstract}
We perform nuclear magnetic resonance (NMR) measurements of the oxygen-17 Knight shifts for \sro, while subjected to uniaxial stress applied along [100] direction. The resulting strain is associated with a strong variation of the temperature and magnetic field dependence of the inferred  magnetic response. A quasi-particle description based on density-functional theory calculations, supplemented by many-body renormalizations, is found to reproduce our experimental results, and highlights the key role of a van-Hove singularity. The Fermi liquid coherence scale is shown to be tunable by strain, and driven to low values as the associated Lifshitz transition is approached.
\end{abstract}

\maketitle

\sro\ is widely recognized as the paradigmatic example of a very clean, strongly correlated Fermi Liquid (FL) with a simple quasi-two-dimensional Fermi surface (FS), from which an unconventional superconducting ground state emerges~\cite{Mackenzie2003,Georges2013}. While the superconducting state has been a subject of intense study~\cite{Mackenzie2017,Mackenzie2020,Kivelson2020}, there are multiple attributes of \sro\ that motivate an in-depth analysis of the normal state, including the ability to characterize the FL state and quasiparticle dispersions with exquisite accuracy and the opportunity to study how the quasiparticles gradually lose coherence as temperature $T$ is raised, evolving all the way into `bad metal' behaviour at high-$T$~\cite{Tyler1998,Mravlje2011,Kugler2020}. 

Indeed, FL behavior applies only below a characteristic crossover
temperature $\TFL\sim 30$~K~\cite{Maeno1997} and is characterized,
among other properties, by the expected thermal variation of the
resistivity $\delta\rho\sim T^2$ and an enhanced $T$-independent Pauli
susceptibility~\cite{Maeno1994,Imai1998}. Interestingly, before
settling into the Fermi liquid temperature independent susceptibility,
the NMR signal displays a shallow maximum at about 40K~\cite{Imai1998}.

In the low-$T$ FL regime, highly accurate determinations of the quasiparticle properties have been achieved using quantum oscillations (QO)~\cite{Bergemann2003,Mackenzie1996}, angle-resolved photoemission spectroscopy (ARPES)~\cite{Damascelli2000,Shen2007,Tamai2019}, and optical conductivity measurements~\cite{Stricker2014}. The fermiology consists of three quasiparticle bands and associated FS sheets, conventionally labelled $\alpha$, $\beta$, $\gamma$,  forming predominantly from a hybridization of each of the Ru $d_{xy}$, $d_{zx,yz}$ orbitals with oxygen 2$p$ orbitals. 
Spin-orbit coupling (SOC)
affects the orbital character of
the quasiparticles states 
and leads to a high 
degree of orbital mixing for the $\beta$ and $\gamma$ branches along the Brillouin 
Zone diagonal~\cite{Haverkort2008,Veenstra2014,Zhang2016,Kim2018,Tamai2019}. 
The gradual breakdown of the quasiparticle picture at higher temperatures is revealed by transport measurements (resistivity)~\cite{Tyler1998} and Hall effect~\cite{Shirakawa1995,Mackenzie1996a,Zingl2019} as well as ARPES and optical spectroscopy~\cite{Stricker2014}. 

The bands are strongly renormalized~\cite{Mackenzie1996} by electronic correlations.  
Theoretical work suggested~\cite{Mravlje2011} that these correlations result from the combined effect of (i) the Hund's rule coupling~\cite{Georges2013,Luo2019}, which has led to a characterization of \sro as a member of the broad family of `Hund's metals' and (ii) importantly, the proximity of the Fermi level to a van-Hove singularity (vHs) associated with the quasi-2D $\gamma$ band~\cite{Kugler2020}. 
The effects of passing $E_F$ through the vHs were experimentally studied with doping~\cite{Shen2007}, straining of thin films~\cite{Burganov2016} and application of uniaxial stress~\cite{Barber2018,Sunko2019}. This last technique allows the effects of the vHs to be probed without introducing additional disorder. This was exploited also in a NMR study that revealed enhancements of the Knight shifts at the critical strain~\cite{Luo2019}. Theoretically, the study of the temperature dependence of the NMR response was reported~\cite{Mravlje2011} for the unstrained case, {\color{black}but the detailed response of the Fermi liquid correlations to applied stress is still an open question. The ARPES results reported in Refs.~\cite{Shen2007,Burganov2016} found no evidence for increased mass enhancement near the vHs in the former, and only a weak enhancement in the latter. On the other hand, a significant increase of the $T^2$ coefficient associated with resistivity was reported under applied uniaxial stress~\cite{Barber2018}. This was later found to be consistent with a Boltzmann description of transport based on coherent quasiparticles close to a van Hove singularity~\cite{Herman2019, Stangier2021}.}

{\color{black}Here we address the role of the vHs by acquiring \O\ NMR data in the crucial temperature range around $\TFL\sim 30$~K, under conditions of variable uniaxial stress. The results provide direct evidence that the crossover scale $\TFL$ is controlled by the location of the vHs relative to Fermi level $E_F$, and moreover that the associated singular DOS strongly influences the physical properties such as spin susceptibility over an extraordinarily broad temperature range. }
$\TFL$ is driven to {\color{black}a vanishingly small value} at the critical strain, and the spin susceptibility 
 inferred from the \O\ Knight shift measurements exhibits the expected logarithmic temperature dependence for a two-dimensional vHs. 
 The non-FL magnetic response at the critical strain is marked by a strong nonlinear field dependence~\cite{Luo2019}, owing to comparable Zeeman and thermal energy scales, and a corresponding divergent singularity at the chemical potential.
 Our experimental results are shown to be in good agreement with a theoretical analysis based 
 on a quasiparticle description starting from the band-structure evaluated under strain but keeping quasiparticle renormalizations independent on strain. This agreement 
 hence provides further support for the limited role of strain on the quasiparticle renormalizations in this material. 

\section*{Results}

\subsection*{NMR experiments under uniaxial strain}

To study the Fermi-liquid crossover upon approaching a van Hove singularity, we performed \O\ NMR experiments on \sro\ under in-plane uniaxial stress $\varepsilon_{aa}$ in a temperature range 1.5--50~K at applied field strengths $B=3\mathrm{ T}$ and $8\mathrm{ T}$.
The magnetic field $\mathbf{B}\parallel b$ results in NMR intensity from three oxygen sites, labeled here 
as two in-plane sites O(1), O(1$'$), and apical site O(2). For the O(1) site the neighbouring Ru sites are along $b$ direction, parallel to the magnetic field, hence the corresponding Knight shift is labeled $K_{1 \parallel}$ and for O(1'), where the neighbouring Ru sites are  perpendicular to the magnetic field, the Knight shift is labeled as $K_{1' \perp}$. (The in-plane site geometry is defined in Fig.~\ref{fig:expGeometry} in the Supplementary Material.) Correspondingly, the hyperfine couplings are different, leading to distinct NMR absorption frequencies even in absence of strain. Taking into account also the electric quadrupolar coupling of the five $I=5/2$ \O\ transitions, results in 15 total NMR absorption lines. In this work we focus on the NMR shift of the central transitions for the O(1) and O(1') sites, which are rendered crystallographically inequivalent due to the $B_{1g}$ component of the strain.  As such, the strain-dependent quadrupolar effects~\cite{Luo2019} are subtracted out in the analysis, so as to isolate the hyperfine contribution to the total shift. 

\begin{figure}[ptb]
\centering
\includegraphics[width=1\columnwidth]{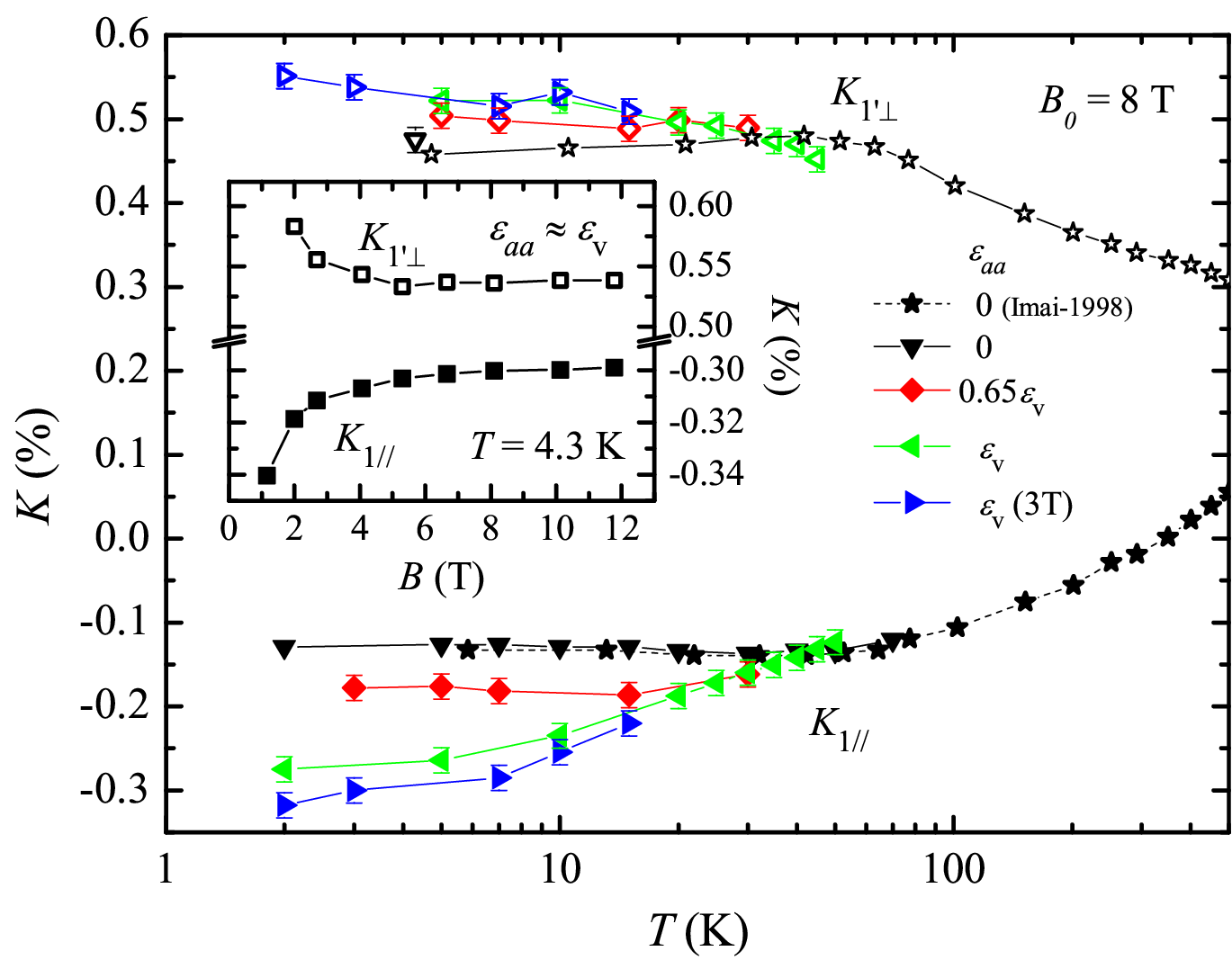}
\caption{Temperature-dependent \O\ NMR Knight shifts, for different fields and strains. Note that for the critical strain, the effects of the singularity are more substantial in $K_{1\parallel}$ than in $K_{1'\perp}$ Inset: {\color{black} Shift vs field data from \cite{Luo2019}.} The shifts at the critical strain $\varepsilon_v$ exhibit a strong field dependence, observable for sufficiently low temperatures, that is associated with the Zeeman split of the singularity. 
}
\label{shifts-all}
\end{figure}

The measured shifts are shown in Fig.~\ref{shifts-all} where strain is seen to have a pronounced effect on the temperature dependence of the normal state behavior, and particularly so for the O(1) site. In the unstrained case (black), an extremum is seen in the data at $\sim 40$K  followed by a crossover to the $T$-independent shift expected for a FL for temperatures $T<\TFL\sim$30 K~\cite{Imai1998,nourafkan18}.  $\TFL$ thus obtained is consistent with the FL crossover temperature observed by other methods~\cite{Mackenzie2003}. The crossover temperature is observed to shift to lower value upon application of the \a-axis stress ($\varepsilon_{aa}=0.65\varepsilon_v$, red).  At the critical strain ($\varepsilon_{aa}=\varepsilon_v$, green), $\TFL\to0$. Additionally, at sufficiently low temperature the magnetic response is distinctly nonlinear at $\varepsilon_v$, shown in the inset of Fig.~\ref{shifts-all}.


{\color{black}Both} the strong temperature dependence of the Knight shift under unstrained conditions, and the striking low-temperature variations of the shifts, especially in $K_{1\parallel}$ while subjected to strain can be interpreted in the band-structure framework in terms of proximity to the van-Hove singularity in the $\gamma$ band. For strains lower than the critical one, the van-Hove singularity (as shown in Fig.~\ref{fig:DMFT-DOS}) is located at positive energy $E_{\mathrm{vHs}}$. The non-monotonic dependence of the Knight shift emerges due to a thermal depopulation of states as temperature drops below $E_{\mathrm{vHs}}$. Under strained conditions, the Fermi level moves toward the energy at the vHS, which explains both higher values of the Knight shift and the vanishing of the crossover scale.


\subsection*{Theoretical modeling of quasiparticle response}
To make this discussion more quantitative,  we consider a simple theoretical modeling  
in terms of quasiparticles. We introduce the quasiparticle Hamiltonian:  
$\HQP\,=\,\sqrt{\hat{Z}}\,\HKS\,\sqrt{\hat{Z}}$. 
In this expression, $\HKS$ is the Hamiltonian of Kohn-Sham states obtained from density-functional theory (DFT) 
(we use the generalized gradient approximation), and $\hat{Z}$ is a matrix of quasiparticle weights which reflect the correlation-induced renormalizations relating physical electrons to low-energy quasiparticles. 
We construct $\HKS$ by performing DFT calculations for a set of strains between $0.0$\% and $0.8$\%. We apply strain in the $\langle 100\rangle$ direction and scale the $\langle 010\rangle$ and $\langle 001\rangle$ direction according to the experimentally determined Poisson ratios, $-\varepsilon_{yy}$ / $\varepsilon_{xx}$ = 0.508 and $-\varepsilon_{zz}$ / $\varepsilon_{xx}$ = 0.163~\cite{Barber2019}.
We consider a minimal set of three low-energy bands, construct maximally localized Wannier functions~\cite{MLWF1,MLWF2} of t$_{2g}$ symmetry and express $\HKS$ in this minimal basis set of localized orbitals. The matrix of quasiparticle weights $\hat{Z}$ is diagonal in this localized basis and we use $Z_{xy}=0.166$ and $Z_{xz/yz}=0.275$ for $T\le 30$~K. These values are the renormalisations found from DMFT calculations in the Fermi-liquid regime of the unstrained material~\cite{Kugler2020}, which are consistent with the mass enhancements obtained from ARPES~\cite{Tamai2019}, optical spectroscopy~\cite{Stricker2014} and quantum oscillations experiments~\cite{Mackenzie1998}. We also take into account that, above $\sim 30$~K, the renormalisation is gradually reduced as temperature is increased, using Z(T) obtained in Ref.~\cite{Kugler2020}. 

\begin{figure}[tb]
\centering
\includegraphics[width=\columnwidth]{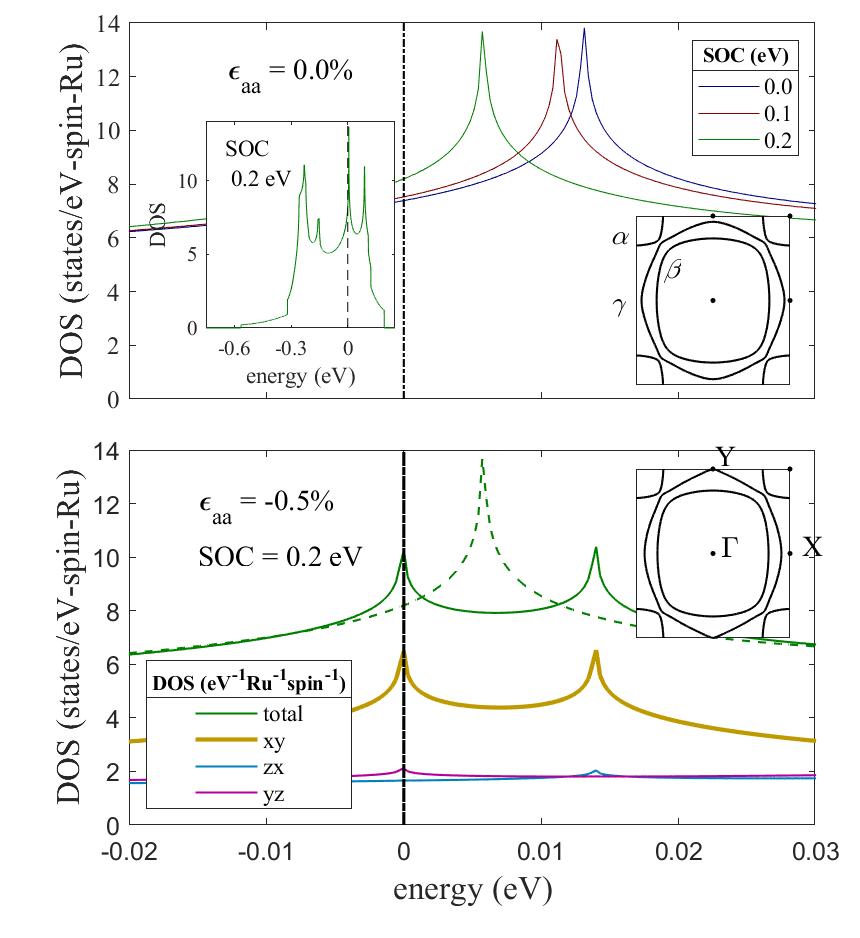}
\caption{\textit{Top:} The $t_{2g}$ quasiparticle density of states (QDOS) as defined in the text.  The main panel displays the region around $E_F$ while illustrating the influence of the spin-orbit coupling on the location of the van-Hove singularity. 
The inset shows the QDOS in a broader energy window. The Fermi surface is also shown (right inset).
\textit{Bottom:} Decomposition of the QDOS into contributions from the $xy$, $zx$, $yz$ orbitals, at the 
critical strain at which the Fermi energy coincides with the van-Hove singularity. The inset displays the corresponding Fermi surface.
}
\label{fig:DMFT-DOS}
\end{figure}

Spin-orbit coupling (SOC) plays a key role in the physics of \sro. As in Refs.~\cite{Zhang2016,Kim2018,Tamai2019,Linden2020,Karp2020}, we take SOC into account in $\HKS$ as a local atomic term in the $t_{2g}$ basis. The bare DFT value of the SOC is about \SI{0.1}{eV}. However, it is established from both theoretical calculations and experiments~\cite{Zhang2016,Kim2018,Tamai2019,Linden2020} 
that electronic correlations lead to an effective enhancement of the SOC to roughly \SI{0.2}{eV}. 
The quasiparticle density of states (QDOS) associated with $\HQP$ for unstrained \sro\ for 0.0, 0.1 and 0.2 eV SOC is shown in Fig.~\ref{fig:DMFT-DOS}(a). We see that including the (enhanced) SOC moves the vHs closer to the Fermi level. As a consequence, the critical strain corresponding to the Lifshitz transition is strongly reduced, an observation also made in Refs.~\cite{Barber2019,Luo2019}. 
By taking the enhancement of the SOC into account, we find a critical strain $\varepsilon_{aa}\sim$-0.5\%, as shown in Fig.~\ref{fig:DMFT-DOS}(b), which is in good agreement with the experimental value of -0.44$\pm$0.06\%~\cite{Barber2019}. We note that by taking only the bare SOC into account, the critical strain is nearly a factor of two too large. 

\begin{figure}[tb]
\centering
\includegraphics[width=\columnwidth]{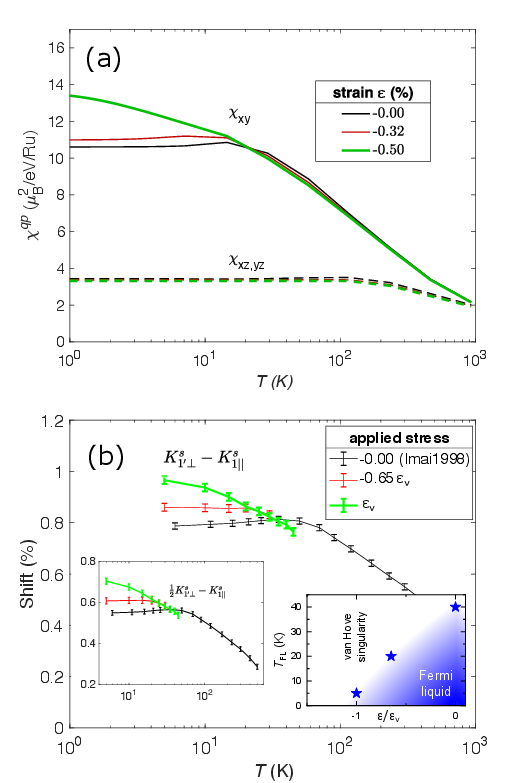}
\caption{(a)  Susceptibility evaluated from the quasiparticle Hamiltonian as a function of temperature and a-axis strain, $\chiqp(\varepsilon_{aa},T)$. The susceptibility is resolved into contributions from the three $t_{2g}$ orbitals. The critical  (compressive) strain is close to $\varepsilon_{aa}=-0.50$\%. (b) Experimentally determined $K_{1'\perp}^s-K_{1||}^s$ as a function of temperature and applied stress, referenced to the critical stress $\varepsilon_v$. $K^s$ denotes the spin part of the total shift. To calculate the difference continuously, shifts have been interpolated from Fig~\ref{shifts-all} after subtraction of the known orbital contribution. ($K_{1||}^o = +0.18\%$, $K_{1'\perp}^o = 0.0\%$)~\cite{Ishida1998}. The inset on the right illustrates the Fermi-liquid crossover temperature as a function of uniaxial strain, extracted from our Knight-shift data. The defining criterion is the shift extremum, which separates the regimes of $T$-independent behavior, and that of logarithmic $T$-dependence. } 
\label{fig:DFTchi_orbital}
\end{figure}

%
We calculate the magnetic susceptibility $\chiqp(T)$ associated with the non-interacting quasiparticle Hamiltonian $\HQP$ by adding a magnetic field term and placing ourselves in the linear response regime (for details, see \ref{sec:appA} of the Supplementary Material). 
Orbitally resolved results are displayed in Fig.~\ref{fig:DFTchi_orbital}, as a function of temperature and for several values of the strain. The $xz,yz$ component has a weak temperature and strain dependence, corresponding to a featureless shape of the density-of-states for that orbital. In contrast, $\chiqp_{xy}$ depends strongly on temperature and strain, due to the proximity to a 2D van-Hove singularity, 
resembling the behavior observed for the Knight shifts. 
For all values of the strain except the critical one, $\chiqp_{xy}$ displays a temperature-independent (Pauli) plateau at low-$T$. 
The temperature below which the plateau behaviour is found decreases as strain is increased, and vanishes at the critical strain. 
On warming to temperatures $T>\TFL$, the susceptibility first increases, as expected from the fact that a large density of states is thermally accessible close to the vHs, and then decreases on further warming. {\color{black}Secondary to the effect of the vHs, the temperature dependent quasiparticle renormalization also contributes to the decrease in $\chi^{qp}$ above $30$ K, as shown explicitly in the Supplementary Material.}
%

It is instructive to place our simple description in the broader context of a Landau theory of the Fermi liquid state. 
In such a theory, quasiparticles interact and the magnetic susceptibility (in a simple single-band framework) is given by: $\chi = \chi_0 \frac{m^*/m}{1+F_0^a}$ where $m^*/m$ is the quasiparticle mass renormalisation, $F_0^a$ is a Landau interaction parameter 
and $\chi_0$ is the bare susceptibility in the absence of any renormalisations. 
Our simple model only takes into account the $\chi_0 \frac{m^*}{m}$ part of this expression (note that within a local description of the Fermi liquid state $m^*/m=1/Z$). 
In other words, we assume that the dominant action is in the change in quasiparticle dispersions (fermiology of the system) as strain is applied, and not in the strain dependence of the Landau interaction parameters, which we assume to be weak. 
We also neglected the possible strain-dependence of the effective mass enhancements $\hat{Z}$. 
Our working assumption of strain independent renormalizations {\color{black}is supported by good agreement with the experimental data presented here, but should be further quantified by other means, such as measurements of specific heat~\cite{Li2021} or high-resolution angular-resolved photoemission experiments.}
. 
It should also be addressed by more elaborate computational approaches, 
such as dynamical mean-field theory, when the methods will allow for reliable solutions in the presence of spin-orbit coupling and low temperatures $<10$K - for recent progress in this direction, see Refs.~\cite{Linden2020,Cao2021}. 



\section*{Discussion}
In order to make a semi-quantitative comparison of the calculations to the experiment it is convenient to take linear combinations 
of the measured Knight shifts to extract the orbital contributions to the susceptibility~\cite{Imai1998}. 
Indeed,  the measured $^{17}$O Knight shifts in Fig.~\ref{shifts-all} include contributions from the spin responses associated with each of the three Fermi surfaces, as well as orbital contributions. Previous \O\ NMR work ~\cite{Ishida1998} determined the orbital shift for the two sites to be $K_{1||}^o = +0.18\%$, $K_{1'\perp}^o = 0.0\%$; both values are consistent with recent examinations of the superconducting state~\cite{Pustogow2019,Chronister2020}. Subtracting these terms from the total shifts shown in Fig.~\ref{shifts-all} leaves behind the contributions proportional to the electronic spin response $K_s$. The remaining $^{17}$O shift contributions are known to arise mostly from dipolar coupling to the occupied $p$-orbitals~\cite{Imai1998,Mukuda1998}. The tetragonal geometry then implies a coupling to the magnetization of the in-plane $p$-orbital for the O(1$'$) site that is twice as large and with opposite sign to that of the O(1), while the couplings to the out-of-plane orbitals dominating the $\alpha$, $\beta$ bands are equivalent. As such, subtracting the shifts of both sites, as shown in the main panel of Fig.~\ref{fig:DFTchi_orbital}(b), eliminates the shared contribution from the $d_{xz/yz}$ bands. 
Hence this quantity is, to first approximation, proportional to the $d_{xy}$ susceptibility. (The inset removes the factor 2 weighting for the O(1$'$) shift and is representative of the average over the $\gamma$-band states within $T$ of $\mu$. Such a linear combination does not fully eliminate the contribution from $xz/yz$ states, but due to the weak temperature and strain dependence of  $\chi_{xz/yz}$ this contribution is just an approximately constant offset). 


It should be noted however that under strained conditions the in-plane oxygen $p_x$ and $p_y$ orbitals are no longer equivalent, which
complicates the analysis. The effects of the asymmetry are amplified by the relative sensitivities of the O(1), O(1$'$) sites to the singularity at Y. Namely, the momentum-dependent overlap $|b_\mathbf{k}|$ of oxygen 2$p$ states with the hybridized $\gamma$-band wave functions is far greater for the O(1) site, than for the O(1$'$) site for momenta near Y~\cite{Luo2019}. 
We discuss further this effect in \ref{sec:appB} of the Supplementary Material. 

So, how do the calculated $\chi_{xy}$ and the difference between the two oxygen shifts compare? At zero strain (black points), 
there is good semi-quantitative agreement. 
At $\varepsilon_{aa}$ = $0.65\,\varepsilon_v$ (red), and $\varepsilon_{aa}$ = $\varepsilon_v$ (green), the calculated susceptibility has the same qualitative behavior as the measured $K_{1'\perp}^s-K_{1||}^s$. 
As strain is increased, the FL coherence temperature is driven to zero $\TFL\rightarrow 0$ and $E_{\gamma}(Y)\to0$ ($\varepsilon \rightarrow \varepsilon_v$). 
Furthermore, the agreement between the low temperature strain enhancement of the susceptibility is also reasonable, with about 23\% enhancement observed in $K_{1'\perp}^s-K_{1||}^s$ compared to~ $\simeq17\%$ in the calculated $\chi_{xy}$ at 4K. The calculated unstrained FL crossover temperature in Fig.~\ref{fig:DFTchi_orbital}(a) is slightly lower than that seen in the measured shifts (about 20~K, or reduced by half relative to the maximum at 40 K, see the right inset of Fig.~\ref{fig:DFTchi_orbital}b), a discrepancy that can be explained by the underestimation of $E_F-E_{vHs}\simeq $~7~meV inherent to the calculation (experiments suggest $E_F-E_{vHs}\simeq $~10-14~meV)~\cite{Shen2007}. As such, the ability of the quasiparticle framework described in this work to reproduce the salient behavior observed in the measured $^{17}$O Knight shifts provides strong support for the interpretation that the Fermi liquid crossover, as well as the approximately logarithmic $T$-dependence for $T>\TFL$, is a consequence of the close proximity of $E_F$ to a (quasi-)2D singularity in the DOS.Interestingly, judging from the agreement, the quasiparticle renormalizations do not substantially increase under strain at least in the studied temperature range. For a system at the actual 2D singularity one could expect the mass enhancement to also exhibit a logarithmic growth upon lowering $T$ and whether this growth simply occurs with a small prefactor and hence quantitatively the effects are small down to 2K, or,  alternatively, other effects such as small but finite warping of the Fermi surfaces in the $z$ direction could also play a role remains to be investigated in future work.

{\color{black} In considering the implications for the superconducting ground state, it is natural to expect that the strain response of $T_c$ might similarly be dominated by the QDOS enhancement. For example, in a BCS weak coupling theory we have $T_c=\omega e^{-1/N_{qp}V}$, with the coupling constant $V$ and $N_{qp}$ the QDOS. Using the results calculated here, $\delta N_{qp}/N_{qp}\simeq15\%$, leads to $\delta T_c/T_c=2.5$. (An unusual feature of the present case is that the strongly field-dependent shifts at the critical strain imply that the QDOS varies substantially over the gap scale.) Note that, by symmetry, the gap vanishes at Y for any odd parity order parameter, hence the above scenario is consistent with the prior evidence for even parity superconductivity from Knight shift measurements~\cite{Pustogow2019,Chronister2020}}  

We emphasize that, in order to keep our theoretical description simple, we did not consider 
two possible effects which should be addressed in future work.
The first is a possible strain-dependence of the quasiparticle mass enhancements, and the second is the strain-dependence of the Stoner enhancement factor $1/(1+F_0^a)$. 
{\color{black}However, the striking agreement with experiment found here, neglecting these effects, suggests they are not crucial to understanding the Fermi liquid behavior near the Lifshitz transition.} 
We note however that an increase of the Stoner factor under strain was found in a previous density-functional theory calculation~\cite{Luo2019}. 
Another interesting issue is the departure from Fermi liquid theory at the critical strain~\cite{Luo2019,Stangier2021}.

To summarize, the normal state transition from an incoherent-like regime to a Fermi liquid-like regime in \sro\ is shown tunable by the application of in-plane strain. The crossover temperature is driven to lower values as the critical strain is approached, as shown in the right inset of Fig.~\ref{fig:DFTchi_orbital}(b), corresponding to the Lifshitz transition for \sro. The results imply that the proximity to the van-Hove singularity is a dominant factor for the macroscopic normal state properties. {\color{black} In spite of the strong effect of the van-Hove singularity on the crossover temperature, which is consistent with what is found in transport~\cite{Barber2019,Herman2019}, the effects of the strain on the quasiparticle renormalization are found to be limited. In comparison to earlier work on doped samples or films~\cite{Shen2007,Burganov2016} the level of disorder is smaller and the singularity correspondingly sharper: the insensitivity of renormalizations to strain is a surprising result that calls for further investigations.} 


\acknowledgments We thank Steve Kivelson and Igor Mazin for helpful discussions. A.~P. acknowledges support by the Alexander von Humboldt Foundation through the Feodor Lynen Fellowship. A.~C. acknowledges support from the Julian Schwinger Foundation. This work was supported by the National Science Foundation under grant numbers 1709304, 2004553. Work at Los Alamos was supported by the Los Alamos National Laboratory LDRD Program. N.~K. is supported by a KAKENHI Grants-in-Aids for Scientific Research (Grant Nos. 17H06136, 18K04715, and 21H01033), and Core-to-Core Program (No. JPJSCCA20170002) from the Japan Society for the Promotion of Science and by a JST-Mirai Program grant (No. JPMJMI18A3). J.~M. acknowledges funding by the Slovenian Research Agency (ARRS) under Program
No. P1-0044, J1-1696, and J1-2458. The
work at Dresden was funded by the Deutsche Forschungsgemeinschaft -
TRR 288 - 422213477 (projects A10 and B01). The Flatiron Institute is a division of the Simons Foundation.

\section*{Methods}
\subsection*{Experimental Details}
An \O\ enriched single crystal of \sro with dimensions 3.0 mm x 0.4 mm x 0.2 mm was mounted onto a piezo-electric variable stress/strain device~\cite{Razorbill} such that the applied uniaxial stress was aligned with the crystallographic \a-axis. In this device, the rectangular bar is clamped on two ends, such that the strained portion of approximately 1 mm length forms the bridge between the clamps. The subsequently wound inductive coil and resonant tank circuit (10-50 MHz) was configured for top-tuning/matching, such that the stress axis coincides with the coil symmetry axis, and the field orientation \BBzero$\parallel$\b\ when placed in the variable temperature cryostat. Compressive stress was applied to the sample \textit{in-situ} via an external voltage to the piezo-electric stacks of the strain device. The NMR frequencies of the \O\ central transitions ($-1/2\leftrightarrow+1/2$) were measured at applied field strengths $B=3\mathrm{ T}, 8\mathrm{ T}$, strains ($\varepsilon_{aa}=0$, $ \varepsilon_{aa}=0.65\varepsilon_v$, $\varepsilon_{aa}=\varepsilon_v$), and covering temperatures $T=1.5-50$ K.For the O(1$'$) site there is a discrepancy between the absolute values of the unstrained shift reported in \cite{Imai1998} compared to those of \cite{Mukuda1998}, \cite{Luo2019}, and the current work. However, the temperature dependent changes which are of interest here are consistent throughout. As such, to compare appropriately a constant offset of $-0.055\%$ was added to the $K_{1'\perp}$ data of \cite{Imai1998} plotted in Fig. \ref{shifts-all} and Fig. \ref{fig:DFTchi_orbital}.  (Throughout, the strains are referenced to the critical compressive strain at the Lifshitz transition, estimated as $\varepsilon_v=\varepsilon_{aa}=-0.44$\%~\cite{Barber2019}).

\subsection*{Computational Details}
\label{sec:appA}

We use the experimental lattice parameters a = b =  \SI{3.8613}{\angstrom},
c = \SI{12.7218}{\angstrom} from Ref.~\cite{Vogt1995} measured at \SI{8}{K}. The internal coordinates
(apical O and Sr z-position) of the unstrained structure were relaxed within DFT.
resulting in z$_O$ =  0.16344 and z$_{Sr}$ = 0.35230 (compared to the experimental
ones of: 0.1634(4) and 0.3529(4))  For simplicity
we do not optimize the internal coordinates under strain. We confirmed with
calculations at higher strains that the internal coordinates are not affected
much by the applied strain. The DFT calculations have been performed with
WIEN2k~\cite{Blaha2018} within the generalized gradient approximation~\cite{PBE}, using RKmax = 8 and a shifted 27 x 27 x 27 k-grid.

For the Wannier construction we use wien2wannier~\cite{wien2wannier} and
Wannier90~\cite{wannier90}, a 10 x 10 x 10 k-grid and a frozen energy window ranging from -1.77 to 3 eV for all strains.

For the calculation of the susceptibility we evaluate the up/down spin DOS under an applied magnetic field of \SI{2}{T}
on a very dense $3360 \times 3360 \times 3360$ k-grid and an energy grid with a spacing of \SI{0.28}{meV}.

Parts of the calculations have been performed using the TRIQS library~\cite{TRIQS}.

\section{Data Availability}
Excel data will be deposited in https://www.pa.ucla.edu/content/sr2ruo4-knightshift-vs-temperature.

\normalem
\bibliography{FLcrossover3_bib}

\section{Author Contributions}
A.C., A.P., {\color{black}Y.L.}, and S.E.B. designed experiments; A.C.,  A.P. and {\color{black}Y.L.} performed experiments; A.C., A.P., and S.E.B. analyzed data; M.Z., J.M., and A.G. designed computations; M.Z. performed computations; N.K., D.A.S., F.J., and E.D.B. contributed new reagents/analytic tools; N.K. monitored sample growth; D.A.S., F.J., C.W.H., and A.P.M. contributed to sample characterization; E.D.B. contributed to sample alignment, cutting, and oxygenation; and A.C., M.Z., A.P., C.W.H., A.P.M., J.M., A.G. and S.E.B. wrote the manuscript.
\section{Competing Interests}
The Authors declare no Competing Financial or Non-Financial Interests.

\newpage

\appendix

\section*{Supplementary Information}
\subsection*{Ru $d$-O 2$p$ hybridization and the hyperfine fields at the $^{17}$O sites}
\label{sec:appB}

\begin{figure}[htp]
\centering
\includegraphics[width=1\columnwidth]{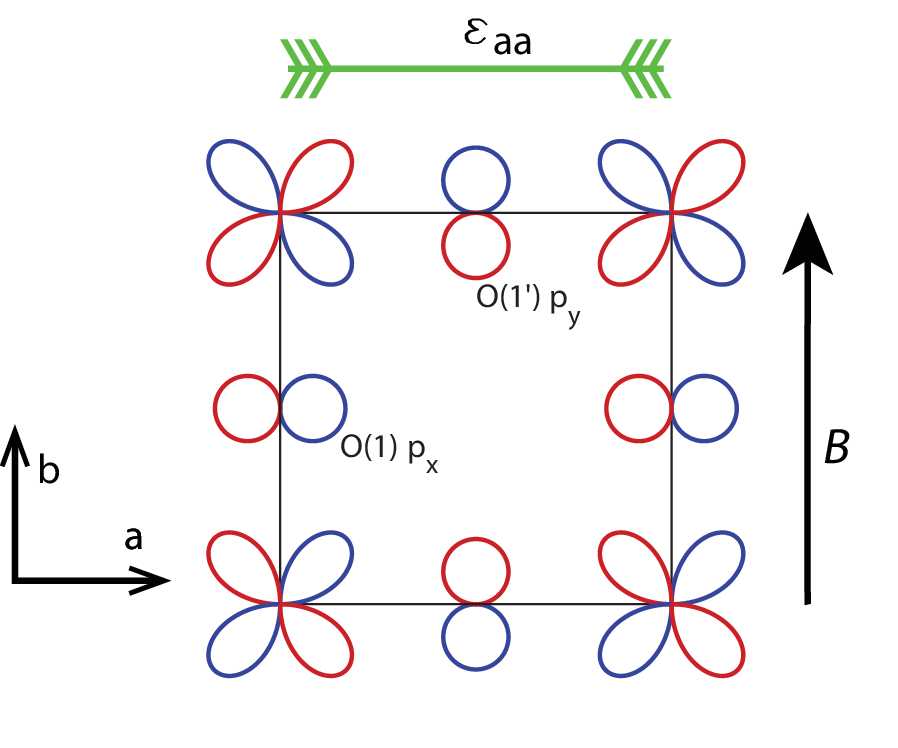}
\caption{Predominant hybridizing orbitals of the quasi-2D $\gamma$ band. The relative phases at the \textbf{Y} point of the Brillouin Zone are indicated by the coloring.}
\label{fig:expGeometry}
\end{figure}

The bands crossing the Fermi surface are comprised of hybridizing Ru 4$d$ $t_{2g}$ and O 2$p$ orbitals. Shown in Fig.~\ref{fig:expGeometry}, $d_{xy}-p_{x}$ ($d_{xy}-p_{y}$), correspond to the predominant orbital character of the $\gamma$-band state, and therefore are the most relevant here. Particularly consequential to the presented shift data is the dipolar $^{17}$O nuclear spin coupling to the oxygen 2$p_x$ ($p_y$). (Not shown are the $d_{xz}-p_z$ and $d_{yz}-p_z$ orbitals which extend out-of-page and predominate the character of the quasi-1D $\alpha$, $\beta$ bands.)

The applied uniaxial stress breaks the symmetry of O(1), O(1$'$) sites. In particular, the relative shift sensitivity of the two sites to the vHs at \textbf{Y} is very different, with O(1) much more sensitive than O(1$'$). The mismatch is linked to the greater overlap of O(1) $p_x$ with the hybridized states in the vicinity of the vHs. A clear signature of the asymmetry at \textbf{Y} {\color{black}$(\mathbf{k}=[0,\pi])$} is highlighted by the orbital coloring in Fig.~\ref{fig:expGeometry}, which illustrates that {\color{black}the states at $Y$ include hybridization to the O(1) $p_x$ orbital (the phases of the adjacent Ru $d_{xy}$ lobes are the same) but not to the O(1') $p_y$ orbital (the corresponding phases are opposite).}  

The interaction can be represented by a diagonal 3$\times$3 matrix, with elements $A(-1,-1,2)$, with
\begin{equation}
    A\equiv\frac{2}{5}\Braket{r_p^{-3}}.
\end{equation}
Using $1/\Braket{r_p^3}=3.72/a_0^3$~\cite{Mukuda1998} ($a_0$ the Bohr radius), $A\sim91~\textrm{kOe}/\mu_B$~\cite{Mukuda1998}. In the specific experimental geometry illustrated in Fig.~\ref{fig:expGeometry}, $A_{d\parallel}(1)=-A, A_{d\perp}(1')=2A$.

The significance of the O(1,1$'$) site inequivalence is clear after considering the projection of single-particle states near to the Fermi energy into the $p_x$ and O(1$'$) $p_y$ orbitals. To evaluate such a dipolar contribution to the Knight shift $K_d(1,1')$, the usual approach results in $K_{di}=A_{ii}\chi^s$. This result is quantitatively insufficient for uniaxially stressed \sro, since it does not take into account the momentum dependence of the wavefunction overlap $b_\mathbf{k}\equiv\Braket{O_{2p}|u_\mathbf{k}}$. Specifically, $|b(1)_\mathbf{k}|^2\gg|b(1')_\mathbf{k}|^2$ for $\mathbf{k}$ near to the density of states singularity at \textbf{Y}. Then, the expression for the dipolar contribution to the shifts is
\begin{align}\label{eq:DipoleShift}
    K_{d\parallel}&=\frac{-Ag\mu_B}{2B}\int dE_\mathbf{k}dA_\mathbf{k}|b(1)_k|^2\\
    &\times[g(E_{\mathbf{k}\downarrow},A_\mathbf{k})f(E_{\mathbf{k}\downarrow})\nonumber\\
    &-g(E_{\mathbf{k}\uparrow},A_\mathbf{k})f(E_{\mathbf{k}\uparrow})],\nonumber\\
    K_{d\perp}&=\frac{2Ag\mu_B}{2B}\int dE_\mathbf{k}dA_\mathbf{k}|b(1')_k|^2\\
    &\times[g(E_{\mathbf{k}\downarrow},A_\mathbf{k})f(E_{\mathbf{k}\downarrow})\nonumber\\
    &-g(E_{\mathbf{k}\uparrow},A_\mathbf{k})f(E_{\mathbf{k}\uparrow})],\nonumber
\end{align}
In the expressions above, $g(E_\mathbf{k}\uparrow,A_\mathbf{k})$ is the density of states per unit energy-Fermi surface area, and $f(E_\mathbf{k},\uparrow)$ is the Fermi function for state $\mathbf{k}$, kinetic energy $E_\mathbf{k}$. The corresponding magnetization is written
\begin{align}
    M=\mu_B\int dE_\mathbf{k}dA_\mathbf{k}\\
    &\times[g(E_{\mathbf{k}\downarrow},A_\mathbf{k})f(E_{\mathbf{k}\downarrow})\\
    &-g(E_{\mathbf{k}\uparrow},A_\mathbf{k})f(E_{\mathbf{k}\uparrow})],\nonumber
\end{align}
These expressions lead to $T$-dependent susceptibilities for comparable thermal and $E_v^\gamma$ energies. And further, if the Fermi energy is proximate to the vHs, a nonlinear bulk magnetization in the case of comparable Zeeman, thermal and  $E_v^\gamma$ energy scales, 

Finally, note that more typically, the factors $|b_\mathbf{k}|^2$ in the expressions for shift and magnetization are taken as constant, then folded in with the coupling constant $A$. The density of states factors are also assumed constant, with the end result that $K\sim\chi$.

\subsection*{Influence of the temperature dependence of renormalization}
\begin{figure}[htp]
\centering
\includegraphics[width=1\columnwidth]{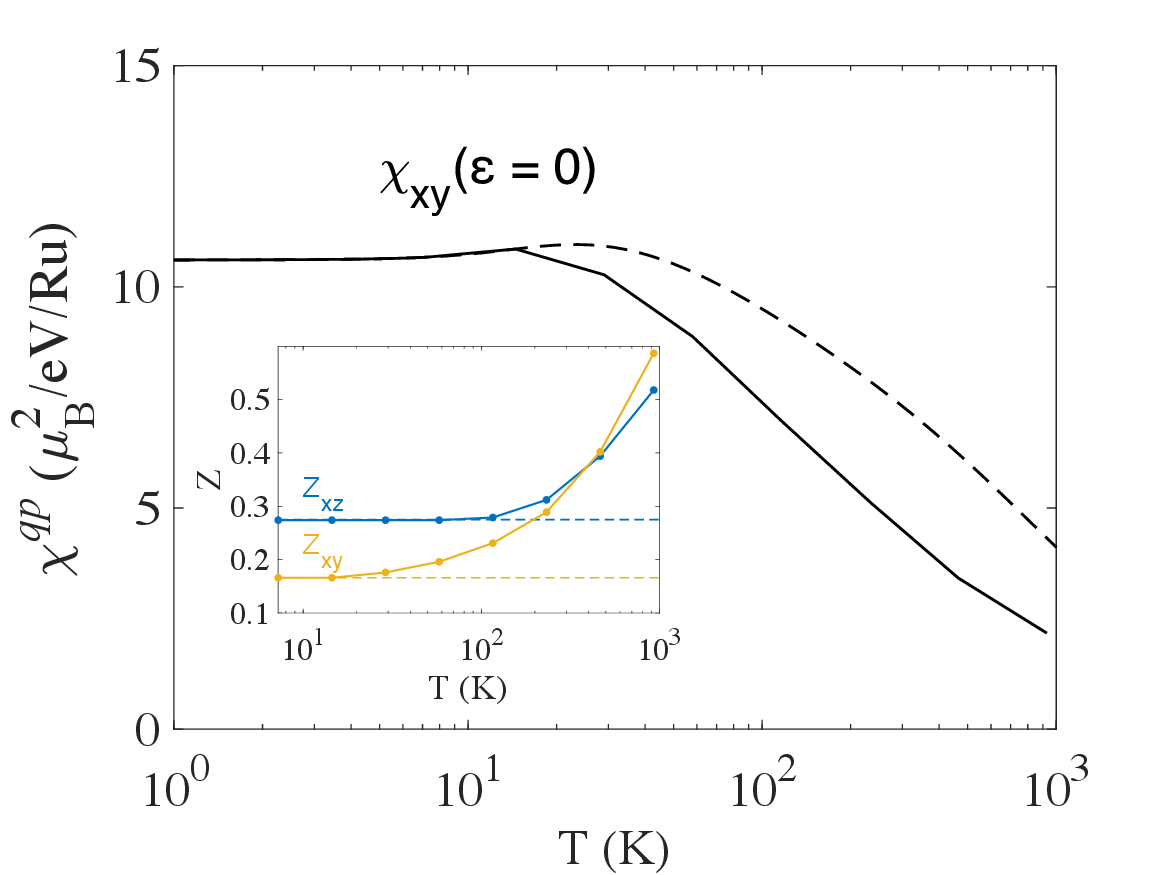}
\caption{Calculated $\chi_{xy}$ using the temperature dependent renormalizations from~\cite{Kugler2020} as in the text (solid line) compared to that using a constant $Z_{xy}=0.166$ (dashed line). Inset: Z(T), reproduced from Ref.~\cite{Kugler2020}.}
\label{fig:ZvTeffect}
\end{figure}

{\color{black}
Fig.~\ref{fig:ZvTeffect} illustrates the role  of the temperature dependence of quasiparticle renormalization $Z(T)$. The quasiparticle susceptibility calculated with the effect of $Z(T)$ included is shown with full line whereas the dashed line depicts the result when $Z$ in the calculations is fixed to its zero-temperature value.
}

\end{document}